\documentclass[%
reprint,
 amsmath,amssymb,
 aps,
 prl]{revtex4-2}

\usepackage{graphicx}% Include figure files
\usepackage{dcolumn}% Align table columns on decimal point
\usepackage{bm}% bold math
\usepackage[many]{tcolorbox}
\usepackage{lipsum}
\usepackage{amsmath,amssymb,cancel}

\usepackage{shuffle}
\usepackage{graphbox}
\usepackage{environ}
\usepackage{physics}
\usepackage{tabularx}
\usetikzlibrary{shapes.geometric}
\usepackage{asymptote}
\usepackage{slashed}
\usepackage{subfig}
\usepackage{float}
\usepackage{cleveref}

\newcommand*{\shifttext}[2]{%
\settowidth{\@tempdima}{#2}%
\makebox[\@tempdima]{\hspace*{#1}#2}%
}

\newcommand{\fwl}{\mathrm{F.L.}}

\begin{document}

\title{All-loop planar integrands in Yang-Mills theory from recursion}

\author{Qu Cao$^{1,2}$}
\email{qucao@zju.edu.cn}
\author{Fan Zhu$^{2,3,4}$}
\email{zhufan22@mails.ucas.ac.cn}
\affiliation{
$^{1}$Zhejiang Institute of Modern Physics, Department of Physics, Zhejiang University, Hangzhou 310027, China\\
$^{2}$CAS Key Laboratory of Theoretical Physics, Institute of Theoretical Physics, Chinese Academy of Sciences, Beijing 100190, China \\
$^{3}$School of Fundamental Physics and Mathematical Sciences, Hangzhou Institute for Advanced Study, UCAS \& ICTP-AP, Hangzhou, 310024, China\\
$^{4}$School of Physical Sciences, University of Chinese Academy of Sciences, No.19A Yuquan Road, Beijing 100049, China
}
\date{\today}

\begin{abstract}
In this letter, we  generalize the recursion methods based on cut equations~\cite{Arkani-Hamed:2024pzc}, originally developed for scalar theories, to gluons in pure Yang-Mills theory. In gauge theories, planar loop integrands are subtle to defined and obtained due to  the existence of scaleless integrals. A critical challenge arises when constructing higher-loop integrands via recursion methods: integrands with or without all scaleless terms both are incompatible with the reconstruction formalism in~\cite{Arkani-Hamed:2024pzc}. To address this, we introduce a {\it refined} integrand by systematically removing specific scaleless contributions, and develop an algorithmic implementation of the recursion to all-loop level. We explicitly demonstrate the framework by three steps and obtain the recursion formula in pure Yang-Mills theory. In the ancillary files, we provide the results up to the two-loop five-point integrand and the simplified result in the large-$D$ limit for the three-loop four-point case.
\end{abstract}

\maketitle

%\tableofcontents

\section{Introduction}
The study of loop integrands in gauge theories has seen significant progress in supersymmetric theories~\cite{Arkani-Hamed:2010zjl,Arkani-Hamed:2010pyv,Geyer:2015bja}. Ironically, much less progress has been made in pure Yang-Mills (YM) theory in generic dimensions, particularly at higher loops. At tree level, string-inspired and worldsheet formalisms provide a clear framework for YM theory~\cite{Cachazo:2013gna,Cachazo:2013iea,Cachazo:2013hca,Mason:2013sva}, and the gravity amplitudes can be obtained via the double-copy method~\cite{Bern:2008qj,Bern:2010ue,Bern:2019prr}. Beyond one loop, fewer formalisms have been developed~\cite{Bern:1990cu,Bern:1990ux,Bern:1991aq,Geyer:2015jch,Geyer:2017ela}. Recently, one of the authors and collaborators proposed the scaffolded-gluon program~\cite{Arkani-Hamed:2023jry} based on surfaceology~\cite{Arkani-Hamed:2023lbd,Arkani-Hamed:2023mvg}, which successfully constructed ``the'' one-loop integrand~\cite{Arkani-Hamed:2024tzl}. Nonetheless, challenges remain at higher loops, despite some progress with formal techniques, such as the "Q-cut'' formalism~\cite{Baadsgaard:2015twa} and the two-loop integrand from Double-Forward limit~\cite{Geyer:2019hnn}.These methods, while conceptually valuable, remain difficult to implement for obtaining explicit results. The current frontier of computation is the two-loop four-point integrand, obtained through bootstrapping the numerators under the color-kinematics duality~\cite{Bern:2013yya,Li:2023akg}.  This raises a crucial question: How can higher-loop integrands be computed more efficiently and accurately? Inspired by the cut equations in scalar theories~\cite{Arkani-Hamed:2024pzc}, we investigate how recursion relations can be applied to gluon loop integrands via the single-cut approach. At one loop, this recursion works and naturally expands into a sum of gauge-invariant building blocks~\cite{Cao:2024olg}. However, at two loops, complications arise, causing the naive recursion to break down. In this letter, we discuss how to resolve these issues and extend the recursion to higher-loop gluon integrands.

The scalar loop integrands are simple, featuring constant numerators in their expressions. We review the recursion for $\text{tr}(\phi^3)$ scalar theory in the appendix.  However, when extending the recursion to the gluon case with nontrivial numerators, several issues arise, particularly those related to scaleless integrals~\footnote{A \textit{scaleless integral} is an integral that has no dependence on kinematic invariants and  vanish under a regularization scheme, such as dimensional regularization. }.  

The main issues are as follows: 
1. \textbf{Incorrect residues from missing scaleless terms}: 
Scaleless contributes to integrands, though not integrals. A critical issue is that certain scaleless terms at $(L{-}1)$-loop can become non-scaleless at $L$-loop due to the forward limit. The recursion fails because the integrand develops incorrect residues.
2. \textbf{Obstruction from scaleless integrals}: A more serious issue arises when retaining all scaleless terms, such as the massless bubble variables $X_{i,i+1}$ in the numerators. When performing a single cut on the (L{-}1)-loop integrand, these unwanted $X_{i,i+1}$ terms appear in the sum over polarization states and may cancel poles in the integrand, causing extra contributions to the residue (we will discuss this in detail below). 
3. \textbf{Symmetry deformation of the residues}: The desired integrand should  possess the cyclic symmetry of the external points and the permutation symmetry of the internal loop punctures. As upon issues mentioned, the certain residues of some non-scaleless terms are scaleless, which we find out that break the symmetry between residues that inherited from the integrand. Hence, the residues obtained form the forward limit are {\it deformed} in scaleless terms. Fortunately, we can still construct the integrand with the desired symmetries from deformed residues.

From these observations, we conclude that a proper integrand should retain certain scaleless terms to ensure a correct single cut, which is necessary to maintain the validity of the recursion at any loop order.  

In this letter, we identify and classify scaleless integrals based on planar variables/Feynman diagrams. This helps us define a \textit{refined} integrand $\hat{I}_{n}^{L}$ for the $L$-loop  $n$-point pure Yang-Mills theory in a generic dimension $D$, which has removed specific scaleless integrals. Using the refined integrand, we obtain a recursion~\eqref{eq: reconstruct from dres} for all loop integrands of pure Yang-Mills theory based on the single cut, which from the forward limit (F.L.).

Our recursion procedure is summarized as follows:
\begin{equation}\label{eq: procedure}
	\hat{I}_{n+2}^{L-1}\xrightarrow[\text{step 1}]{\mathrm{F.L.}}
    \delta_n^L+\underset{\scalebox{0.7}{$X_{i,z_L}=0$}}{\mathrm{dRes}}\hat{I}_{n}^{L}
    \xrightarrow[\text{step 2}]{\mathrm{Red.}}
	\underset{\scalebox{0.7}{$X_{i,z_L}=0$}}{\mathrm{dRes}}\hat{I}_{n}^{L}\xrightarrow[\text{step 3}]{\mathrm{Sym.}}\hat{I}_{n}^{L}\,,
\end{equation}
where $L\geq1,\,n\geq4$~\footnote{Note that because of the scaleless terms, we only focus on the physical integrands at $n\geq4$, (all integrals at $n\leq3$ are scaleless).} and $\delta_n^L$ denotes the collection of all unwanted scaleless terms generated from Step 1, the \textit{Forward Limit} (F.L.)~\footnote{When discussing whether the terms in the residue are scaleless, we assume that the pole is restored.}. The $\underset{\scalebox{0.7}{$X_{i,z_L}=0$}}{\mathrm{dRes}}\hat{I}_{n}^{L}$ denotes the deformed residue, which differs from $\underset{\scalebox{0.7}{$X_{i,z_L}=0$}}{\mathrm{Res}}\hat{I}_{n}^{L}$ only in some specific scaleless integrals. Step 2, \textit{Reduction (Red.)}, removes \(\delta_n^L\) to obtain the deformed residues. Finally, Step 3, \textit{Symmetrization } (Sym.), restores \(\hat{I}_{n}^{L}\) from the deformed residues by imposing its intrinsic symmetry (cyclic group $\times$ permutation group).

\section{Recursion from single cuts}
\subsection{Planar variables and \textbf{Forward-Limit (F.L.)}.}
We focus on the loop integrands in the large $N_{c}$ limit, with the gauge group SU$(N_{c})$, in pure Yang-Mills theory. The leading $N_{c}$ terms are the so-called $\textit{planar integrand}$, which can be embedded in the planar diagrams. 
\begin{figure}[H]
	\centering
	\subfloat[]{\includegraphics[scale=0.45]{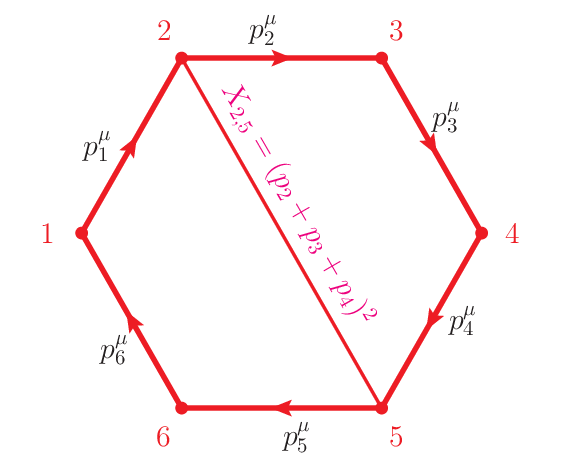}\label{fig: planarvar1}}
    \subfloat[]{\includegraphics[scale=0.45]{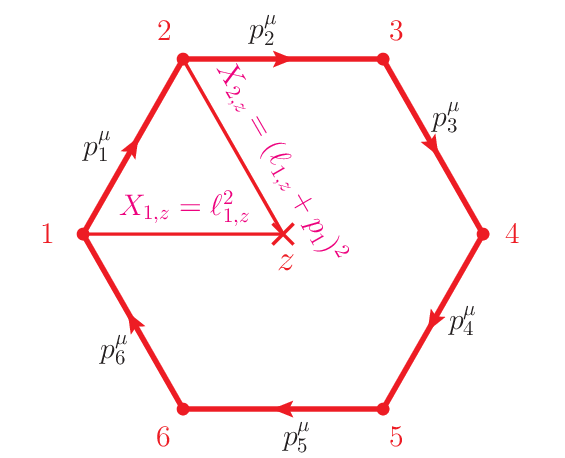}\label{fig: planarvar2}}\hfill
	\caption{The example of a hexagon on the surface. Determining planar variables of curves on the surface via homology, at tree-level (left) and one-loop (right).}
	\label{Fig: planarvar}
\end{figure}
For planar diagrams, we can use the dual diagrams to manifest the symmetry of loop momentum $\ell\to\ell+k$, with arbitrary momentum $k$. All propagators/denominators are associated with the so-called planar variables $X_{i,j}$. One of our basic starting points is to transform all the scalar products $k\cdot k$ and $\ell\cdot k$ into planar variables. Let us have a quick review on planar variables. In the dual diagram, the curve $\mathcal{C}_{a,b}$ that ends with two dual points $a,b$ on the surface is assigned the planar variables $X_{a,b}$ (see~\Cref{Fig: planarvar})~\cite{Arkani-Hamed:2023lbd}. The $L$-loop $n$-point planar integrand $I_{n}^{L}$ relates to the $n$-gon with $L$ punctures inserted, we refer to the dual points $\{1,2,\ldots,n\}$ and $\{z_1,z_2,\ldots,z_L\}$ as marked points and loop punctures, respectively.

The relations between momentum $k_i^\mu/\ell_{i,z_j}^\mu$ and the dual points are, which originally from the ABHY relations~\cite{Arkani-Hamed:2017mur}:
\begin{equation}\label{eq: dualpoint&kinematic}
\begin{aligned}
	&k_i^\mu= (x_{i+1}-x_i)^\mu\equiv (x_{i+1,i})^\mu\,,\\
    &\ell_{i,z_j}^\mu= (x_{i}-x_{z_j})^\mu\equiv (x_{i,z_j})^\mu\,,\\
	&(x_{a,b})^\mu (x_{c,d})_\mu\equiv\frac{1}{2}\left
    (X_{a,d}+X_{b,c}-X_{a,c}-X_{b,d}\right)\,,
\end{aligned}
\end{equation}
where all the $X_{a,b}=X_{b,a}$. We call the planar variables like $X_{i,i},X_{i,i+1},X_{z_i,z_i}$ that correspond to boundary curves as boundary planar variables (which can be regarded as “regulators” for massless bubble and tadpole diagrams). There are $\frac{n(n-3)+L(L-1)}{2}+nL$ of $X_{a,b}$'s (excluding boundary planar variables), which is exactly the number of independent scalar products. For the Lorentz products of loop momenta and polarizations, we adopt the basis set $\{\ell_{i,z_j}\cdot\epsilon_i|1\leq i\leq n,\,1\leq j\leq L\}$.
 
The $\hat{I}_{n}^{(L)}$ can be regarded as a meromorphic function, where the denominator consists of $X$ terms, while the numerator contains $X$, $\ell\cdot\epsilon$, $k\cdot \epsilon$ and $\epsilon\cdot\epsilon$ terms. The (deformed) residues of $\hat{I}_{n}^{(L)}$ play a crucial role in reconstructing the integrand. From the cut equations, we focus on the residue at $X_{i,z_{L}}$, which corresponds to the so-called \textit{forward limit} of the $I_{n+2}^{(L-1)}$ integrand, where two adjacent particles are glued together with opposite momenta~\cite{Simon_TreeLoop}.
\begin{figure}[H]
	\centering
	\includegraphics[scale=0.3]{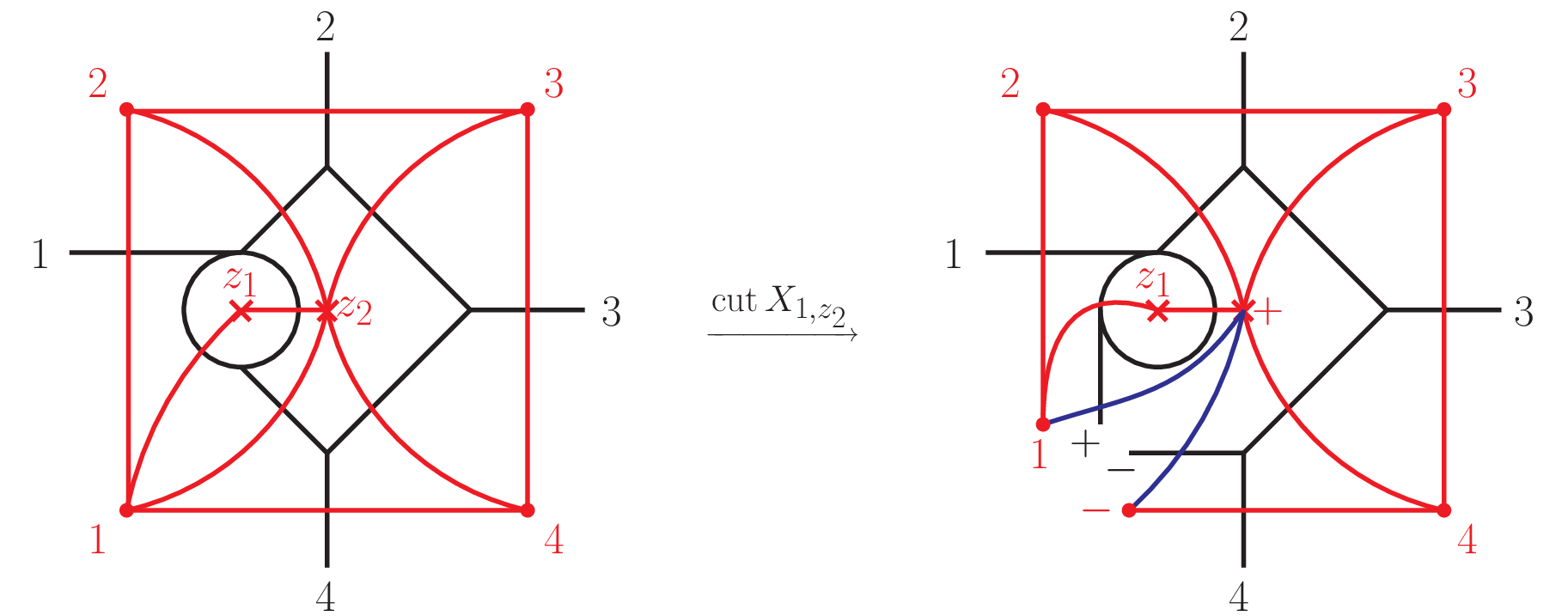}
	\caption{\small{Illustration of the cut on $X_{1,z_2}$ of one 4-point 2-loop diagram.
    }}
	\label{Fig: cutexp1}
\end{figure}

The forward limit of the YM integrand $\hat{I}_{n+2}^{L-1}$ can be divided into two parts. For those planar variables $X$'s and momentum $k$'s, there is a map of the dual points on the surface.
\begin{equation}\label{eq: fwl_scalar}
    \begin{pmatrix}
		+&-\\
        \downarrow&\downarrow\\
		z_L&1
	\end{pmatrix}\hat{I}_{n+2}^{L-1}(+,1,\ldots,n,-;\,z_1,\ldots,z_{L-1})\,.
\end{equation}
For example, the momentum $k_+^\mu=(x_i-x_+)^\mu$ is mapped to $(x_i-x_{z_L})^\mu=\ell_{1,z_L}^\mu$, and a graphic illustration of~\eqref{eq: fwl_scalar} shown in~\Cref{Fig: cutexp1}.

For other variables, the extra maps of the forward limit for Lorentz products with polarizations are
\begin{equation}\label{eq: fwl_ym}
	\left\{\begin{aligned}
        &\ell_{1,z_L}\cdot \epsilon_{\pm}\xrightarrow{\fwl} 0\,,\\
		&\epsilon_{-} \cdot \epsilon_+ \xrightarrow{\fwl} D-2\,, \\
		&\epsilon_{-}\cdot u\, \epsilon_{+}\cdot v\xrightarrow{\fwl} u\cdot v\,,
	\end{aligned}\right.
\end{equation}
where $u,v$ represent either momenta or polarization vectors. The first line in~\eqref{eq: fwl_ym} corresponds to the transversality condition, $k_{\mp}\cdot \epsilon_{\pm}=0$. The second and third lines arise from summing over the states of two gluon polarizations, $\sum_{\text{states}} \epsilon_{-}^{\mu}\epsilon_{+}^{\nu}=\eta^{\mu\nu}-\frac{\ell_{i}^{\mu}q^{\nu}+\ell_{i}^{\nu}q^{\mu}}{\ell_{i}\cdot q}$
as derived in~\cite{Kosmopoulos:2020pcd}. 

The equation~\eqref{eq: fwl_scalar} suggest that applying the forward limit inevitably generates certain scaleless terms (\textit{e.g.}, the pole $X_{1,-}$ mapped to $X_{1,1}$, indicating a tadpole diagram). These observations underscore the necessity of dealing with the residue in each recursion step, as prescribed in~\eqref{eq: procedure}.

All the aforementioned issues arise from gluing the polarization vectors of two adjacent gluons~\eqref{eq: fwl_ym}.  In the scalar theory, the residue of the integrand consists only of contributions from $X$ terms~\eqref{eq: fwl_scalar}, which explains why the all-loop recursion works naturally~\cite{Arkani-Hamed:2024pzc}; see also the appendix.  To address these issues, we introduce the \textit{refined} integrand $\hat{I}_{n}^{L}$ from the original integrand $I_{n}^L $ and, in the next subsection, outline the key step of \textit{reduction} its residue using the properties of scaleless integrals.

\subsection{The refined integrand and \textbf{Reduction (Red.)}}
As mentioned earlier, in our recursion for pure YM theory, most scaleless integrals break the symmetry between the residues and lead to significant redundancy. To save the symmetry and kill the redundancy, we define the operation \textit{Reduction (Red.)} as the removal of all terms which never contribute to non-scaleless terms under F.L. in our recursion. The integrand obtained after applying \textit{Red.} is referred to as the \textit{refined} integrand $\hat{I}_{n}^{L}$. 

\begin{figure}[H]
	\centering
    \hspace{1.5em}
	\subfloat[]{\includegraphics[height=0.18\textwidth]{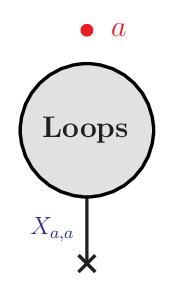}\label{fig: scaleless_sub_diag1}}\hspace{6em}
    \subfloat[]{\includegraphics[height=0.18\textwidth]{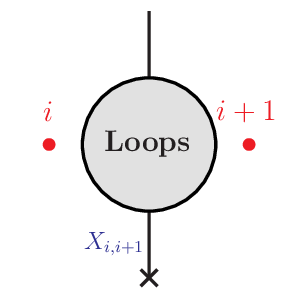}\label{fig: scaleless_sub_diag1}}\hfill

	\subfloat[]{\includegraphics[height=0.12\textwidth]{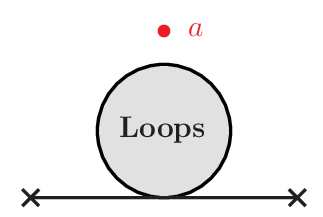}\label{fig: scaleless_sub_diag3}}\hspace{4em}
    \subfloat[]{\includegraphics[height=0.12\textwidth]{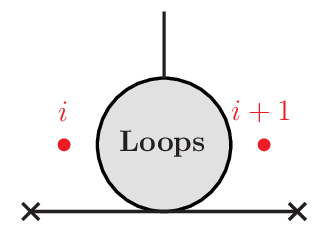}\label{fig: scaleless_sub_diag4}}\hfill
	\caption{Four types of scaleless sub diagrams, where red points are dual points on the surface. The black cross denotes the other remaining parts of the diagram. (a) the tadpole type with $X_{a,a}$ pole, for $a\in\{1,\ldots,n,z_1,\ldots,z_L\}$; (b) the massless bubble type with $X_{i,i+1}$ pole, for $i\in\{1,\ldots,n\}$; (c) the reduced tadpole type; (d) the reduced massless bubble type.}
	\label{Fig: scaleless_sub_duagram}
\end{figure}

We classify the scaleless terms into four types based on their poles and identify which types should be reduced. The four types are: (a) the tadpole type, which involves boundary planar variables $X_{i,i},X_{z_i,z_i}$; (b) the massless bubble type which involves boundary planar variables $X_{i,i+1}$; (c) reduced tadpole type, derived from the tadpole type by canceling the $X_{i,i}$/$X_{z_{i},z_{i}}$ pole; and (d) reduced massless bubble type, derived from the massless bubble type by canceling the $X_{i,i+1}$ pole. In view of Feynman diagrams, all four types of scaleless poles correspond to four types of subsectors (see~\Cref{Fig: scaleless_sub_duagram}), and the reduced tadpole(massless bubble) type can be seen as the subdiagrams of tadpole(massless bubble) type. 

From the mapping~\eqref{eq: fwl_scalar} one can find that the scaleless poles of type (a), (b) and (c) remain scaleless throughout the recursion. 
For types (a) and (c), the tadpole topology preserved under the forward limit.
For type (b), the boundary planar variables, expect for $X_{+,1}$ and $X_{-,+}$, are mapped to boundary planar variables $X_{i,i}, X_{i,i+1}$, ensuring that the corresponding poles remain scaleless. Meanwhile, $X_{+,1}$ and $X_{-,+}$ are mapped to $X_{1,z_L}$, meaning that the resulting poles do not contribute to the residue at $X_{1,z_L}=0$.
For type (d), it may contribute to non-scaleless integrals at the next loop level, as shown in~\Cref{Fig: cutexp1}, the right cut diagram is reduced massless bubble type. 
Now we have a precise statement of \textit{Red.} procedure: 
The \textit{Red.} removes all the scaleless terms of types (a), (b) and (c).

In turn, the outcome of cutting a simple pole $X_{i,z_j}$ of a $L$-loop non-scaleless diagram is always a $(L-1)$-loop non-scaleless diagram or a reduced massless bubble type diagram.

In the appendix, we present a systematic method for identifying the two reduced types (c) and (d) based on their poles, rather than resorting to Feynman diagrams. 
It is also important to point out that, although the reduced massless bubble is not always glued as a non-scaleless diagram, it would still contribute to the non-scaleless terms in further recursions up to the maximal loop level with four or five external legs. 

Let's discuss the issue of retaining all scaleless terms. Recall that the forward limit~\eqref{eq: fwl_ym} transform the product $\epsilon_{+}\cdot u\,\epsilon\cdot v$ into $u\cdot v$. The key point here is that when $u,v$ correspond to some $k_i$ or $\ell_{i,z_j}$, they must be expanded again in terms of the planar variable basis using~\eqref{eq: dualpoint&kinematic}. In this process, some boundary variables would appear in the numerator, but their precise treatment remains unclear, \textit{i.e.}, some should be treated as zero, while others should be regarded as “regulators”. Continuing, the cancellation of boundary planar variables between numerator and denominator is not fully understood. Such cancellation would transform some massless bubble type scaleless terms to reduced massless bubble type scaleless terms, which may contribute to non-scaleless at next loop level. Hence, retaining all scaleless terms is highly problematic and the {\it Red.} removes the issues they cause.

Finally, we emphasize that any non-scaleless terms must contain at least two simple $X_{i,z_j}$ poles (a brief discussion is provided in the appendix). This ensures that each physical term can be detected through multiple residues, leading to the recursion formula~\eqref{eq: reconstruct from dres}. 

% \vspace{1em}

\subsection{The symmetry of $\hat{I}_n^L$ and \textbf{Symmetrise (Sym.)}.}
Since any non-scaleless term contains some simple poles $X_{i,z_j}$, obtaining all undeformed residues $\underset{\scalebox{0.7}{$X_{i,z_j}=0$}}{\mathrm{Res}}\hat{I}_{n}^{L}$ provides the complete information of $\hat{I}_n^L$, and the only remaining questions are the overlap between distinct residues. 
%To deal with the overlap, we should preserve the symmetry of integrand. 
The symmetry of refined integrand is the two symmetry groups $Z_n$ and $S_L$ in $\hat{I}_n^L$, which correspond to the cyclic invariance of market points and the permutation invariance of loop punctures, respectively. These two symmetry groups have a natural interpretation on the surface and we have
\begin{equation}\label{eq: symmetry of int}
	\hat{I}_n^L=
	\hat{I}_n^L(\rho_{1},\ldots,\rho_{n};\, z_{\sigma(1)},\ldots,z_{\sigma(L)}),
\end{equation}
where $\rho\in Z_{n}$ and $\sigma\in S_L$. The symmetry~\eqref{eq: symmetry of int} of $\hat{I}_n^L$ induces the symmetry between residues
\begin{equation}\label{eq: symmetry of res}
	\underset{\scalebox{0.7}{$X_{i,z_j}=0$}}{\mathrm{Res}}\hat{I}_{n}^{L}=(1,\cdots,n)^{i}\cdot(z_j,z_L)\cdot\underset{\scalebox{0.7}{$X_{1,z_L}=0$}}{\mathrm{Res}}\hat{I}_{n}^{L},
\end{equation}
where the brackets denote the cyclic permutation operation on the marked points and loop punctures. 

For undeformed residues, we can use~\eqref{eq: symmetry of res} to generate all residues from a single residue. Removing overlap is straightforward, since the number of repetitions equals the number of simple poles $X_{i,z_j}$. We can reconstruct the refined integrand and generalize the recursion from the cut equation~\cite{Arkani-Hamed:2024pzc} as:
\begin{equation}\label{eq: reconstruct I_n^l}
\begin{aligned}
    \hat{I}_{n}^{L}&=\mathcal{W}\left[\sum_{i=1}^n \sum_{j=1}^L\frac{\underset{\scalebox{0.7}{$X_{i,z_j}=0$}}{\mathrm{Res}}\hat{I}_{n}^{L}}{X_{i,z_j}}\right]\\
    &=\mathcal{W}\left[\sum_{i=1}^n \sum_{j=1}^L(1,\ldots,n)^{i}\cdot(z_j,z_q)\cdot\frac{\underset{\scalebox{0.7}{$X_{p,z_q}=0$}}{\mathrm{Res}}\hat{I}_{n}^{L}}{X_{p,z_q}}\right]\,,
\end{aligned}
\end{equation}
where $p,q$ in the second line are arbitrary. 
The $\mathcal{W}$'s function is multiplying a weight $1/L(\Gamma)$ to each term, where $L(\Gamma)$ is the number of simple poles $X_{i,z_j}$ in that term, which avoids the overlaps \footnote{The effect of $\mathcal{W}$'s function is similar to the integral in~\cite{Arkani-Hamed:2024pzc}, see also the equation~\eqref{eq:rec sca} in the appendix.}. For example, $\mathcal{W}\left(\frac{1}{X_{1,z_1}X_{1,z_2}^2,X_{z_1,z_2}X_{3,z_2}}\right)
    =\frac{1}{2 X_{1,z_1}X_{1,z_2}^2,X_{z_1,z_2}X_{3,z_2}}.$
In this way, we can ignore the complexity that the higher power poles $X_{i,z}^m$ brought and use the minimal seed to construct the entire integrand.

Let us return to the deformed residues derived from F.L. and Red. procedures. We find that the symmetry~\eqref{eq: symmetry of res} of the deformed residue is broken among the reduced massless bubble type of scaleless terms, and we cannot obtain other deformed residues from a single deformed residue by simple permutations (while the symmetry of the $z$’s is still preserved as it corresponds to a relabeling), {\it i.e.},
\begin{equation}\label{eq: symbreak of dres}
    \underset{\scalebox{0.7}{$X_{2,z_j}=0$}}{\mathrm{dRes}}\hat{I}_{n}^{L}\neq
    (1,\ldots,n)\cdot\underset{\scalebox{0.7}{$X_{1,z_j}=0$}}{\mathrm{dRes}}\hat{I}_{n}^{L}\,.
\end{equation}
However, we claim that~\eqref{eq: reconstruct I_n^l} still holds for the deformed residues, provided that the undeformed residue is replaced by the deformed residue
\begin{equation}\label{eq: reconstruct from dres}
\begin{aligned}
    \hat{I}_{n}^{L}&=\mathcal{W}\left[\sum_{i=1}^n \sum_{j=1}^L\frac{\underset{\scalebox{0.7}{$X_{i,z_j}=0$}}{\mathrm{dRes}}\hat{I}_{n}^{L}}{X_{i,z_j}}\right]\,,\quad\text{or}\\
    \hat{I}_{n}^{L}&=\mathcal{W}\left[\sum_{i=1}^n \sum_{j=1}^L(1,\ldots,n)^{i}\cdot(z_j,z_q)\cdot\frac{\underset{\scalebox{0.7}{$X_{p,z_q}=0$}}{\mathrm{dRes}}\hat{I}_{n}^{L}}{X_{p,z_q}}\right]\,.
\end{aligned}
\end{equation}
The result of reconstruction~\eqref{eq: reconstruct from dres} possess the symmetry~\eqref{eq: symmetry of int} is obvious, while the equivalence between the first and second lines in~\eqref{eq: reconstruct from dres} is nontrivial due to~\eqref{eq: symbreak of dres}, and we have verified that they are equal for the 2-loop 4-point case. There must be some intriguing cancellations of the terms that break the symmetry~\eqref{eq: symmetry of res}, and we claim that the symmetry breaking is restored under~\eqref{eq: symbreak of dres}, as indicated by the equivalence between the first and second lines in~\eqref{eq: reconstruct from dres}.

\section{Discussion on results}
We apply the recursion~\eqref{eq: reconstruct from dres} and obtain the two-loop refined integrand in pure YM theory up to five points. In the ancillary \texttt{Mathematica} file, we provide a detailed demonstration of the complete procedure for constructing the 2-loop 4-point pure YM integrand from the 1-loop 6-point pure YM integrand\footnote{The tree-level amplitude data can be obtained from the CHY formula, with a publicly available code in~\cite{He:2021lro}. The one-loop integrand data can be found in~\cite{Cao:2024olg}.}. Additionally, we include the new result for the 2-loop 5-point integrand in the file. Our 2-loop 4-point integrand is consistent with the result of~\cite{Li:2023akg}, up to some scaleless terms. Furthermore, we have verified that all spanning cuts of the 2-loop 5-point  integrand are correct.
The mechanism of symmetry restoration from deformed residues remains incompletely understood, and a further 2-loop 6-point(3-loop 4-point) result may provide deeper insights, but we leave this computation for future work.

 We compare our all-loop recursion with other formal methods~\cite{Baadsgaard:2015twa,Geyer:2019hnn}. The main difference is that other methods rely on the residue theorem, but they introduce spurious poles, {\it i.e.}, linear propagators such as $(\ell+p_{1})^2-\ell^2=2\ell\cdot p_{1}$, which must cancel in the final result.  During the calculation, eliminating all spurious poles is computationally expensive. Our recursion is based on the cut equation, where all poles are physical, {\it i.e.}, quadratic propagators such as $(\ell+p_{1})^{2}=X_{2,z_{1}}$\footnote{There are some methods to obtain the quadratic propagator integrand at one loop~\cite{Feng:2022wee,Edison:2022jln,Dong:2023stt}}.  Moreover, the treatment of scaleless integrals differs. Using planar variables, we systematically classify scaleless integrals and efficiently remove unwanted terms. Therefore, our recursion is more efficient and enables us to obtain new two-loop results recursively.

 In order to simplify the integrand and examine the all-loop recursion, we discuss the recursion for the large-$D$ limit in the appendix. The $L$-loop integrand for pure YM theory in a generic dimension $D$ depends on $D$. For a fixed $L$-loop order, we take the $D\to \infty$ limit to extract the leading $D^L$ terms. The leading large-$D$ terms are simpler, only have scalars running in each loop. The $D^L$ component of the refined integrand $\hat{I}_{n}^{(L)}$ can be further simplified to remove all scaleless integrals while still ensuring the validity of the all-loop recursion. In the large-$D$ limit, we verify that the recursion remains valid and provide the result for the 3-loop 4-point integrand in the ancillary files.

\section{Outlook}
Our preliminary studies have opened up several new avenues for further investigation. 

In the recent paper~\cite{Cao:2024olg}, a similar reconstruction for one-loop amplitudes in general gauge theories has been successfully performed. Generalizing our recursion to higher-loop cases in general gauge theories would be both interesting and challenging due to the lack of a comprehensive understanding of loop amplitudes involving external fermions.

Beyond gauge theories, gravity theories may play an even more significant role. It would be particularly interesting to generalize our recursion based on the single cut to non-planar gravity loop integrands. There is evidence suggesting that such a generalization is promising: at one loop, gravity loop integrands can be constructed using the single cut with linear propagators~\cite{He:2016mzd,He:2017spx}. 
Thus, a similar recursion with quadratic propagators should also work at one loop for gravity.

With the data for two-loop Yang-Mills integrands, one possible way to organize their structure is by finding the BCJ numerators~\cite{Bern:2008qj,Bern:2010ue,Bern:2019prr,Bern:2013yya}. For the two-loop four-point case, global BCJ numerators do not exist; only local BCJ numerators are present~\cite{Li:2023akg}. Therefore, it would be valuable to determine new BCJ numerators for the two-loop five-point case using our results. 

Another possible approach to uncovering the structure of higher-loop integrands is to use the large-$D$ expansion. At one loop, the leading $D$ terms and the sub-leading $D^{0}$ terms are related by differential operators~\cite{Cao:2024olg}. It would be highly exciting to discover similar patterns at two loops using our data!

\section*{Acknowledgments}
We thank Jin Dong, and Song He for stimulating discussions and collaborations on related projects, and we thank Yichao Tang for useful comments on the
manuscript.  Q.C. thanks Institute for Advanced Study for hospitality during his visit. The work of Q.C. is supported by the National Natural Science Foundation of China under Grant No. 123B2075.

% \newpage
\bibliographystyle{apsrev4-1}
\bibliography{Refs}% Produces the bibliography via BibTeX.

\newpage

\widetext
\begin{center}
	\textbf{\Large Supplementary Material}
	\end{center}
\section{Recursion on scalar theory}
    
    Let us briefly review the all-loop recursion for the simplest theory $\Tr(\phi^3)$, based on the cut-equations. The$L$-loop  $n$-point loop integrands $M_{n}^{L}$ for the planar color-ordered $\Tr(\phi^3)$ amplitude are summing over all trivalent diagrams with the constant numerators in the momentum space.  The $M_{n}^{L}$ is determined by the $n$-gon with $L$ inner-punctures inserted, labeled as $\{z_{1},\ldots, z_{L}\}$. In the dual space, the trivalent diagrams are full triangulation diagrams for the $n$-gon with $L$ punctures inserted. All propagators/denominators are associated with the so-called planar variables $X_{i,j}$. The $M_{n}^{L}$ can be viewed as a meromorphic function depending on the $X$'s.
\begin{equation}
    M_{n}^{L}(1,\ldots,n;z_{1},\ldots,z_{L})=\sum_{\Gamma} \frac{n_{\Gamma}}{D_{\Gamma}}
\end{equation}
where $\Gamma$ is a triangulation diagram, the denominator $D_{\Gamma}$ is the polynomial of planar variables, with the constant numerator $n_{\Gamma}$.

The cut equations tell us that the residue at a pole $X=0$ captures the boundary behavior of the $n$-gon. There are three types of poles. The first type is $X_{i,j}=0$, whose residue factorizes into the product of two lower-point sub-polygons. The second type is $X_{i,z_{j}}=0$, where the residue corresponds to the forward limit of a single $(n+2)$-gon with $L-1$ punctures inserted. The third type is $X_{z_{i},z_{j}}=0$, whose residue is related to a double color-ordered $n$-gon with $L-2$ punctures and a new sub 2-gon inserted. The third type is more complicated and requires information from multiple color orderings, which we will not consider here. Our recursion focuses on the second type, $X_{i,z_{j}}=0$. By summing over all residues at these poles, one can reconstruct the full integrand.

Based on the single-cut on $X_{i,z_{j}}=0$, the integrand has the recursion~\cite{Arkani-Hamed:2024pzc}:
\begin{equation}\label{eq:rec sca}
    M_{n}^{L}=\int_{0}^{1}\frac{{\rm d}t}{t}\sum_{i=1}^n \sum_{j=1}^L\frac{\underset{\scalebox{0.7}{$X_{i,z_j}=0$}}{\mathrm{Res}}M_{n}^{L}(t)}{X_{i,z_j}(t)}
\end{equation}
where we shift all the $X_{i,z_{j}}\to X_{i,z_{j}}(t)=X_{i,z_{j}}/t$. 
For the simplest case: $1$-loop $2$-pt, there are three terms: bubble diagram $\frac{1}{X_{1,z_{1}}X_{2,z_{1}}}$, and two tadpole diagrams $\frac{1}{X_{1,z_{1}}X_{1,1}}$, $\frac{1}{X_{2,z_{1}}X_{2,2}}$. The recursion acts as 
\begin{equation}
\begin{aligned}
    M_{2}^{1}&=\int_{0}^{1}\frac{{\rm d}t}{t}\left( \frac{(\frac{1}{X_{1,1}}+\frac{1}{X_{2,z_{1}}(t)})}{X_{1,z_1}(t)}+\frac{(\frac{1}{X_{2,2}}+\frac{1}{X_{1,z_{1}}(t)})}{X_{2,z_1}(t)}\right)\\
   &=\frac{1}{X_{1,z_{1}}X_{2,z_{1}}}+\frac{1}{X_{1,z_{1}}X_{1,1}}+\frac{1}{X_{2,z_{1}}X_{2,2}}\,,
\end{aligned}
\end{equation}
where we already used the conclusion that single-cut on $X_{i,z_{1}}=0$ is the forward-limit of the tree-level $4$-point $M_{4}^{0}(1,2,3,4)=\frac{1}{X_{1,3}}+\frac{1}{X_{2,4}}$.

\section{Scaleless integrals classification and identification}\label{sec_Material}
Let's consider a $L$-loop $n$-point planar diagram $\Gamma$ which doesn't involve boundary planar variables, we define
\begin{equation}\label{def: pgamma}
    P_z(\Gamma):=
    \big\{X_{a,b}|\, X_{a,b}\,\text{is a pole of }\Gamma,\,
    \{a,b\}\cap\{z_1,\ldots,z_L\}\neq\varnothing\big\},
\end{equation}

Note that we have identified $X_{a,b}$ and $X_{b,a}$. We also define $A_\Gamma^i:=\{p|\, p\in\{1,\ldots,n\},\,X_{p,z_i}\in P_z(\Gamma)\}$ is the set of marked points that surround the loop puncture $z_i$, and $B_\Gamma^i:=\{p|\, p=z_i\,\|\, p\in\{z_1,\ldots,z_L\},\,X_{p,z_i}\in P_z(\Gamma)\}$ is the set of loop punctures that surround the loop puncture $z_i$ and itself. For example, if $D_\Gamma=X_{1,z_1}X_{1,z_2}^2,X_{z_1,z_2}X_{3,z_2}X_{1,3}$, we have $P_z(\Gamma)=\{X_{1,z_1},X_{1,z_2},X_{z_1,z_2},X_{3,z_2}\}$, $A_\Gamma^2=\{1,3\}$ and $B_\Gamma^2=\{z_1,z_2\}$. 

\begin{figure}[H]
	\centering
	\includegraphics[scale=0.45]{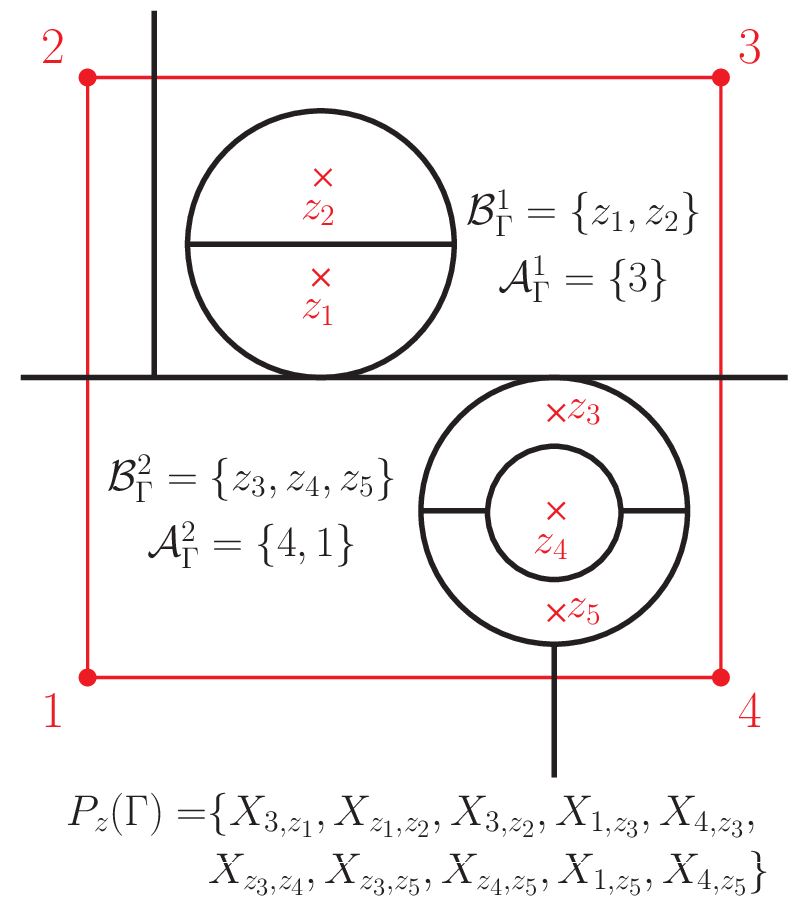}
	\caption{\small{ Example of $P_z(\Gamma)$ for a 4-point 5-loop diagram. We can read $A_\Gamma^i, B_\Gamma^i$ from the $\Gamma$ or $P_z(\Gamma)$, like $A_\Gamma^2=\{3\}, B_\Gamma^3=\{z_3,z_4,z_5\}$. The loop punctures $\{z_1,z_2,z_3,z_4,z_5\}$ distributed into $m=2$ sets, $\mathcal{B}_\Gamma^1=\{z_1,z_2\}$ and $\mathcal{B}_\Gamma^2=\{z_3,z_4,z_5\}$, then we have $\mathcal{A}_\Gamma^1=\{3\}, \mathcal{A}_\Gamma^2=\{4,1\}$, which both identify $\Gamma$ is scaleless.}}
	\label{Fig: PGamma}
\end{figure}

\begin{figure}[H]
	\centering
	\includegraphics[scale=0.3]{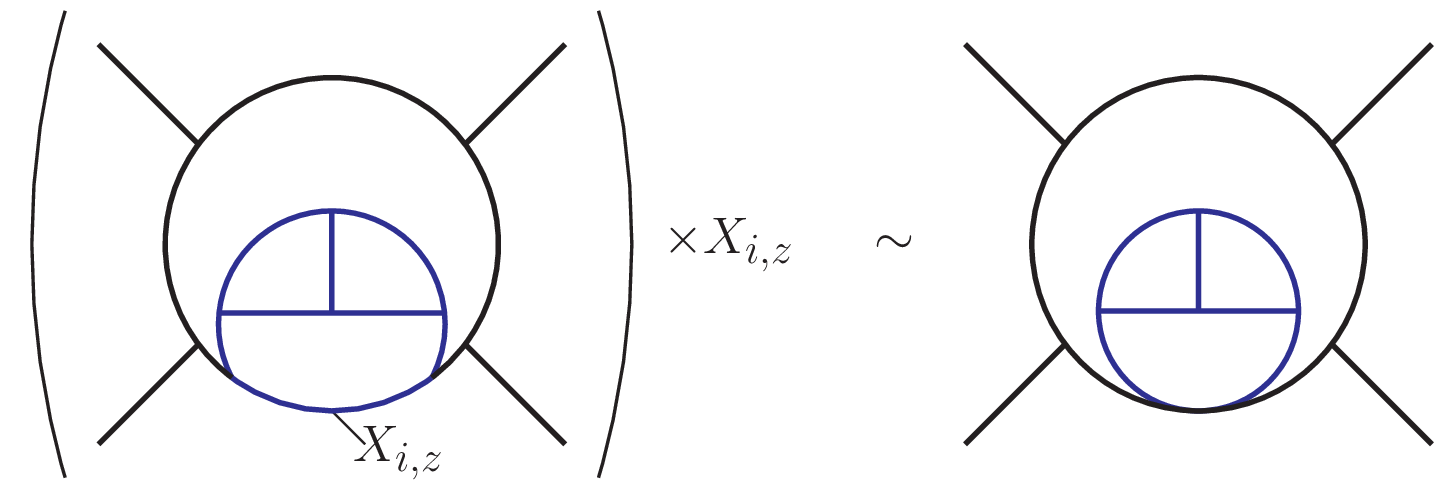}
	\caption{\small{The right diagram contains a  reduced tadpole type diagram which is drawn in blue lines. And if you cut some $X_{i,z}$ of the right diagram you will get a still reduced tadpole type diagram, which means its origin is the left diagram, and the factor $X_{i,z}$ is generated from~\eqref{eq: fwl_ym}}. }
	\label{Fig: tadpole2}
\end{figure}

Then, we distribute the $L$ loop punctures $\{z_1,\ldots,z_L\}$ into $m\,(1\leq m\leq L)$ sets, each set satisfying the definition $\mathcal{B}_\Gamma^i:=\{z_a|\forall z_a\in\mathcal{B}_\Gamma^i, B_\Gamma^a\subseteq \mathcal{B}_\Gamma^i\}$. For each $\mathcal{B}_\Gamma^i$, we define the set $\mathcal{A}_\Gamma^i:=\bigcup_{z_a\in\mathcal{B}_\Gamma^i}A_\Gamma^a$.
%\footnote{In the situation of $m=z_L$, we come to the topology of large-$D$ limit}. 
 The graphic illustrations of $\mathcal{A}_\Gamma^i$ and
$\mathcal{B}_\Gamma^i$ are shown in~\Cref{Fig: PGamma}. If there exist $\mathcal{A}_\Gamma^i$ containing less than two marked points (reduced tadpole type) or two adjacent marked points (reduced massless bubble type), $P_z(\Gamma)$ refers to non-physical pole structure.
There still exist some diagrams of reduced tadpole type that cannot be identified by $\mathcal{A}_\Gamma^i$, and all can be identified as some reduced tadpole type diagrams attached in the inner of a loop on the surface, see to the right of~\Cref{Fig: tadpole2}. 
A further step to identify these diagrams is treating all loop punctures that are surrounded by some marked points as marked points, and then use $\mathcal{B}_\Gamma^i$ and $\mathcal{A}_\Gamma^i$ in the same way as upon. For example, consider a 3-loop 4-point diagram $\Gamma$ with $P_z(\Gamma)=\{X_{1,z_1},X_{2,z_1},X_{4,z_1},X_{z_1,z_2},X_{z_1,z_3},X_{z_2,z_3}\}$, if we treated $z_1$ as a marked point $i_1$, we have $P_z(\Gamma|_{z_1\to i_1})=\{X_{i_1,z_2},X_{i_1,z_3},X_{z_2,z_3}\}$, then we found $\mathcal{B}_\Gamma^1=\{z_2,z_3\}$ and $\mathcal{A}_\Gamma^1=\{i_1\}$, {\it i.e.}, the loops correspond to $z_2,z_3$ form a reduced tadpole type diagram. 

\begin{figure}[H]
	\centering
	\includegraphics[scale=0.5]{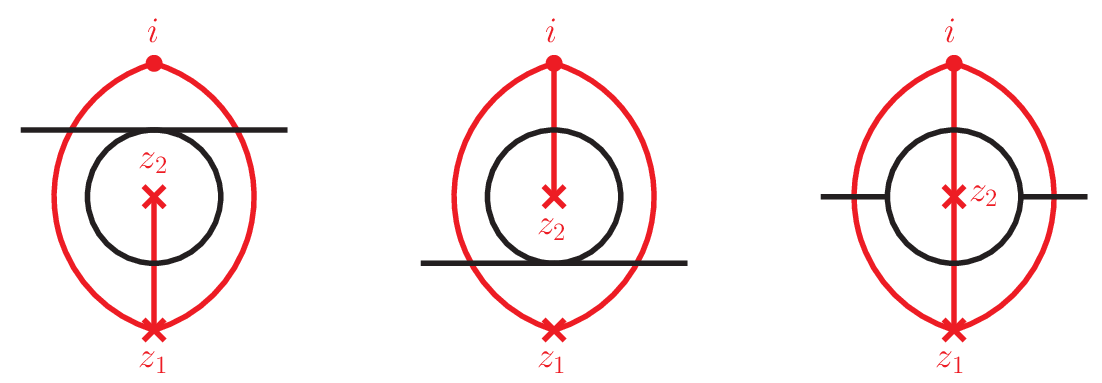}
	\caption{\small{The three types of dual diagrams for double pole $X_{i,z_1}^2$. Only the third case, which contains a simple pole $X_{i,z_2}$, can be regarded as a non-scaleless diagram.}}
	\label{Fig: simplepoleproof}
\end{figure}

There is a brief discussion on the statement ``any non-scaleless
terms must contain at least two simple poles". A stronger version is that ``any higher pole in a non-scaleless diagrams must accompanied by at least one simple pole''. To illustrate this, consider the dual diagram of a double pole $X_{i,z_1}^2$, as shown in~\Cref{Fig: simplepoleproof}. There must be at least one loop puncture inside of the two curves $\mathcal{C}_{i,z_1}$, say $z_2$. It is then straightforward to see that the presence of a simple pole is necessary for the diagram to be non-scaleless.
To analyze the entire diagram rather than just subdiagrams, one can easily observe the presence of at least two simple poles.

\section{Recursion for large-D limit}

In this section, we introduce a limit to simplify the integrand and examine the recursion in this limit. The $L$-loop integrand for pure YM theory in a generic dimension $D$ depends on $D$. For a fixed $L$-loop order, we take the $D\to \infty$ limit to extract the leading $D^L$ terms. The leading large-$D$ terms are simpler, only have scalars running in each loop. The $D^L$ component of the refined integrand $\hat{I}_{n}^{(L)}$ can be further simplified to remove all scaleless integrals while still ensuring the validity of the all-loop recursion.

The large-$D$ limit (or the $D^L$ dependent components) of $L$-loop YM integrands exhibit a well-structured configuration. To establish a recursion for the large-$D$ limit of arbitrary-loop YM integrands, the first modification in Step 1 of~\eqref{eq: procedure} is to replace the operation~\eqref{eq: fwl_ym} by extracting the coefficient of $\epsilon_+\cdot\epsilon_-$ and multiplying it by $(D-2)$ (while still satisfying the transversality conditions) in each recursion cycle.  

\begin{figure}[H]
	\centering
	\includegraphics[scale=0.5]{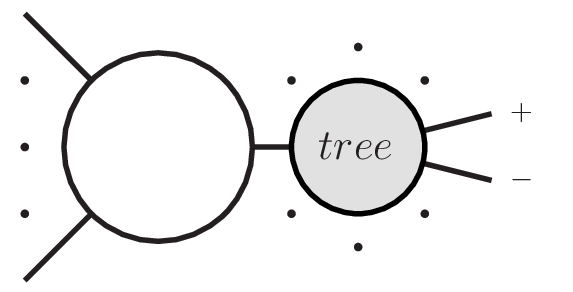}
    %$\propto \epsilon_+\cdot\epsilon_-
	\caption{\small{A one-loop diagram that only scalars running in the loop, {\it i.e.}, the $D$-dependent component of pure Yang-Mills. The legs $+,-$ grow in the same dangling tree. A diagram contributes to the coefficient of $\epsilon_+\cdot\epsilon_-$ only if it admits this configuration. Consequently, the new gluing loop must be apart from the original loop. This induction can extend to higher-loop level.}}
	\label{Fig: largeDtopo}
\end{figure}

With this new operation, one can directly determine the specific topologies of the large-$D$ limit. Notably, each loop is completely decoupled from the others, and $X_{z_i,z_j}$ poles are not allowed on the surface (see~\Cref{Fig: largeDtopo}). This property of topologies forbids the situation of nonphysical topologies gluing as physical topologies. Consequently, in Step 2 of~\eqref{eq: procedure}, we can eliminate all four types of scaleless integrals, rather than only three.

In the large-$D$ limit, we apply the same recursion to obtain integrands up to the $3$-loop $4$-point case and attach the result in the ancillary file.

\end{document}